\documentclass[fp,twocolumn]{jpsj3}

\usepackage{txfonts}
\def\={\hspace{-2mm}&=&\hspace{-2mm}}

\def\etal{{\it et\ al.\ }}

\newcommand{\lsim}
 {\ \raise.35ex\hbox{$<$}\kern-0.75em\lower.5ex\hbox{$\sim$}\ }
\newcommand{\gsim}
 {\ \raise.35ex\hbox{$>$}\kern-0.75em\lower.5ex\hbox{$\sim$}\ }
\def\journal #1#2#3#4{#1 {\bf #2} (#4) #3}
\def\PR{Phys.\ Rev.}

\def\PRB{Phys.\ Rev.\ B}
\def\PRL{Phys.\ Rev.\ Lett.}

\def\JPCS{J.\ Phys.\ Chem.\ Solids}

\def\JPSJ{J.\ Phys.\ Soc.\ Jpn.}

\def\RMP{Rev.\ Mod.\ Phys.}
\def\PTP{Prog.\ Theor.\ Phys.}
\def\ZP{Z.\ Phys.}

\def\EPJB{Eur.\ Phys.\ J.\ B}

\def\NJP{New\ J.\ Phys.}

\def\JPSCP{JPS Conf.\ Proc.}
\def\PP{Phys.\ Proc.}
\title{
Drude and Superconducting Weights and Mott Transitions in 
Variation Theory
} 
 
\author{
Shun \surname{Tamura}
\thanks{E-mail address: shun@cmpt-serv.phys.tohoku.ac.jp} and 
Hisatoshi {\sc Yokoyama} 
}
\inst{
\address{Department of Physics, Tohoku University, Sendai, 980-8578, Japan} 
}

\abst{Drude weight ($D$) is a useful measure to distinguish a metal from an 
insulator. 
However, $D$ has not been justifiably estimated by the variation theory 
for long, since Millis and Coppersmith [Phys.~Rev.~B {\bf 43} (1991) 13770] 
pointed out that a variational wave function $\Psi_Q$, which includes the key 
ingredient (doublon-holon binding effect) for a Mott transition, yields 
a positive $D$ (namely metallic) even in the Mott-insulating regime. 
We argue that, to obtain a correct $D$, an imaginary part must exist 
in the wave function. 
By introducing a configuration-dependent phase factor ${\cal P}_\theta$ to 
$\Psi_Q$, Mott transitions are successfully represented by $D$ 
($D=0$ for $U>U_{\rm c}$) for a normal and $d$-wave pairing states; thereby, 
the problem of Millis and Coppersmith is settled. 
Generally, ${\cal P}_\theta$ plays a pivotal role in describing 
current-carrying states in regimes of Mott physics. 
On the other hand, we show using a perturbation theory, the one-body 
(mean-field) part of the wave function should be complex for band insulators 
such as antiferromagnetic states in hypercubic lattices. 
}

\kword{Drude weight,Variational Monte Carlo Method, Hubbard model, Mott insulator,
band insulator
}
\begin{document}
\maketitle

\section{introduction\label{sec:Intro}} 
The Drude weight $D$, the coefficient of the DC ($\omega=0$) conductivity, 
has been considered to be an important measure to distinguish a metal ($D>0$) 
from an insulator ($D=0$), in particular, since Kohn showed that $D$ can be 
calculated only with quantities with respect to the ground state.\cite{Kohn}  
Actually, $D$ is obtained through $D=d^2 E(A)/dA^2|_{A\rightarrow 0}$ 
($E$: total energy) by introducing a virtual flux (Peierls phase) $A$ or 
equivalently by twisting the boundary condition. 
This formalism is not only convenient for a variety of 
approaches\cite{Shastry,SWZ} but crucial in particular for the variation 
theory. 
\par

More than two decades ago, Millis and Coppersmith\cite{M-C} first applied 
this formalism to variational wave functions for the one-dimensional Hubbard 
model at half filling.\cite{note-dim} 
They used wave functions that include binding factors between a doubly 
occupied site (doublon, D) and an empty site (holon, H),\cite{Kaplan,YS3} 
${\cal P}_Q$, in addition to the usual onsite correlation (Gutzwiller) 
factor,\cite{Gutz} ${\cal P}_{\rm G}$, and showed that this type of wave 
functions ($\Psi_{\rm N}={\cal P}_Q{\cal P}_{\rm G}\Phi_{\rm N}$ with 
$\Phi_{\rm N}$ being a Fermi sea) are metallic even for a sufficiently large 
$U/t$ to be insulating, in the sense that $D>0$. 
At that time, it had not been clarified yet that a D-H factor ${\cal P}_Q$ 
brings about a Mott transition, although the Gutzwiller wave function 
($\Psi_{\rm G}={\cal P}_{\rm G}\Phi_{\rm N}$) was known to be always 
metallic.\cite{YS1,MV} 
Consequently, their result has caused confusion and 
misunderstandings---$\Psi_{\rm N}$ cannot describe a Mott transition--- to 
subsequent studies using $\Psi_{\rm N}$ (and 
$\Psi_d={\cal P}_Q{\cal P}_{\rm G}\Phi_d$ with $\Phi_d$ being a $d$-wave 
BCS state). 
Later, it was confirmed\cite{Y-PTP,YTOT,Capello,YOT,Watanabe,Tocchio} that 
$\Psi_{\rm N}$ ($\Psi_d$) and similar wave functions that have D-H binding 
effects induce Mott transitions at $U_{\rm c}\sim W$ ($W$: band width); 
the behavior in various quantities changes with anomalies at $U_{\rm c}$, 
such as doublon density $d$---an order parameter of Mott transitions---, 
charge-density structure factor $N({\bf q})$, and momentum distribution 
function $n({\bf k})$. 
Thus, the discrepancy between the behaviors of these quantities and $D$ has 
remained an enigma for long years. 
\par

The main purpose of this study is to develop a method for calculating the 
Drude weight appropriately in the variation theory, and settle the above 
problem. 
We first argue that a finite imaginary part is indispensable in the wave 
function for correctly estimating $D$. 
\par

In the regime of Mott physics ($U\gtrsim U_{\rm c}$), we noticed that some 
configuration-dependent phase factor has to be introduced to the ordinary 
(real) wave functions like $\Psi_{\rm N}$ and $\Psi_d$. 
Analyzing the phase added in hopping between a doublon and a holon in the 
field $A$, we construct a phase factor for the present case ${\cal P}_\theta$, 
where $\theta$ is a phase parameter to be optimized. 
For Mott insulators, the optimized ${\cal P}_\theta$ cancels out the Peierls 
phase by satisfying the relation $\theta=A$, and the increment of energy 
owing to $A$ is reduced to zero. 
Thus, $D$ vanishes for $U>U_{\rm c}$ at half filling. 
For $U<U_{\rm c}$ or $\delta>0$ ($\delta$: doping rate), the 
Peierls phase is cancelled out only partially, and $E(A)$ remains larger than 
$E(0)$; $D$ becomes finite. 
In calculations for $A=0$, $\theta$ is optimized at zero
(${\cal P}_\theta=1$), so that the previous results for $d$, $N({\bf q})$ and 
$n({\bf k})$\cite{Y-PTP,YTOT,Capello,YOT,Watanabe,Tocchio} remain unchanging 
for the new wave function $\Psi={\cal P}_\theta\Psi_{\rm N}$. 
Thus, the long-standing issue of Millis-Coppersmith was resolved. 
\par

In fact, we found that this type of configuration-dependent phase factors 
seems generally essential to treat current-carrying states appropriately 
in the regime of Mott physics ($U\gtrsim U_{\rm c}$). 
It was shown that a similar configuration-dependent phase factor is 
indispensable to correctly represent staggered flux states in Hubbard-type 
models.\cite{SF,Toga,d-p}
\par

On the other hand, in treating a band insulator for $U\ll U_{\rm c}$, we 
find ${\cal P}_\theta$ is ineffective. 
For instance, although an antiferromagnetic (AF) state for the square-lattice 
Hubbard model at half filling is insulating for any $U/t$ $(>0)$, 
the Drude weight obtained using 
$\Psi_{\rm AF}={\cal P}_\theta{\cal P}_Q{\cal P}_{\rm G}\Phi_{\rm AF}$
becomes finite for small values of $U/t$. 
It indicates another element is needed for $\Psi_{\rm AF}$. 
We argue using a perturbation theory that the imaginary part in the one-body 
(mean-field) wave function $\Phi_{\rm AF}$ is vital for reducing $D$ in 
this case. 
Here, the imaginary part is given by the first-order perturbation with 
respect to $A$. 
Thus, the mechanism to suppress $D$ is different between Mott and band 
insulators. 
\par

We also address the effectiveness of ${\cal P}_\theta$ for superconducting 
(SC) weight $D_{\rm s}$ in the attractive Hubbard model. 
\par

This paper is organized as follows: 
In Sec.~\ref{sec:formulation}, we introduce trial wave functions for 
estimating the Drude weight through Kohn's formula, and mention why 
$\Psi_Q$ fails in yielding an appropriate $D$. 
In Sec.~\ref{sec:normalSC}, we give the results of VMC calculations using 
${\cal P}_\theta\Psi_Q$ for the normal and $d$-wave pairing states, and 
discuss an improvement of the phase factor. 
In Sec.~\ref{sec:AF}, we consider an AF state as a typical case of 
band insulators of weakly correlations. 
In Sec.~\ref{sec:attractive}, the SC weight is reconsidered for the 
attractive Hubbard model. 
In Sec.~\ref{sec:summary}, we recapitulate the main results in this paper. 
Preliminary results of this study were presented in a 
proceedings.\cite{Tamura-PP}
\par 

\section{formulation\label{sec:formulation}}
%
In Sec.~\ref{sec:model}, we introduce the model and method used in this 
study.  
In Sec.~\ref{sec:wf}, we explain ordinary trial wave functions for $U>0$ 
without a Peierls phase. 
Trial states for $U<0$ will be discussed in Sec.~\ref{sec:attractive}. 
In Sec.~\ref{sec:P-theta}, we show the necessity of a complex wave function 
for a finite Peierls phase, and how to construct a trial wave function 
in regimes of Mott physics. 
In Sec.~\ref{sec:D-and-Ds}, we discuss the relation between Drude ($D$) 
and SC ($D_{\rm s}$) weights in the light of variation theory. 
\par

\subsection{Model and method\label{sec:model}}
%
To study Mott transitions through the Drude weight, we adopt the 
Hubbard model on the square lattice: 
%
\begin{eqnarray}
{\cal H}&=&{\cal H}_{\rm kin}+{\cal H}_{\rm int} \nonumber \\
&=&-t\sum_{\langle i,j\rangle,\sigma}\left(c_{i\sigma}^\dag c_{j\sigma}+
\mathrm{H.c.}\right)+U\sum_j n_{j\uparrow}n_{j\downarrow}, 
\label{eq:H}
\end{eqnarray}
%
where $\langle i,j\rangle$ denotes a sum of nearest-neighbor pairs. 
It is known that, at half filling, the ground state of Eq.~(\ref{eq:H}) 
is insulating with an AF order for any positive $U/t$. 
However, it is important to consider a non-magnetic Mott transition without 
explicitly introducing magnetic orders, because the essence of Mott 
transitions is independent of magnetism; for instance, Mott transitions 
take place in spinless boson systems of ultracold atomic gases.\cite{YMO}
Actually, nonmagnetic Mott transitions were found in the normal and $d$-wave 
pairing branches in the Hubbard model Eq.~(\ref{eq:H}) at $U\sim W$ ($W$: 
band width), and their properties were studied using a few 
methods.\cite{DMFT} 
Furthermore, as we will see later, the AF state also undergoes a crossover 
from a Slater-type (or band) insulator to a Mott insulator at $U\sim W$. 
In this work, we mostly treat repulsive cases, but also study attractive 
cases ($U/t<0$), in which the ground state is $s$-wave SC for any $U/t$, 
but a normal state, as an excited state, undergoes a spin-gap 
transition.\cite{Tamura-attractive} 
\par

To this model, we apply a variational Monte Carlo (VMC) method,\cite{YS1,VMC} 
which yields reliable results for any correlation strength $U/t$ and doping 
rate $\delta=1-N_{\rm e}/N_{\rm s}$ ($N_{\rm e}$: number of 
electrons, $N_{\rm s}$: number of sites) in that the local correlation 
factors are exactly treated.  
As a many-body wave function $\Psi$, the Jastrow type is useful: 
$\Psi={\cal P}\Phi$. 
Here, $\Phi$ indicates a Hartree-Fock-type one-body state, and ${\cal P}$ 
is a product of many-body projection factors. 
In previous VMC 
studies,\cite{Y-PTP,YTOT,Capello,YOT,Watanabe,Tocchio,Miyagawa,doped_Mott} 
Mott transitions were found and the aspects of Mott physics were revealed 
using many-body states $\Psi={\cal P}_Q{\cal P}_{\rm G}\Phi$ or similar 
wave functions.
Note that a D-H binding factor ${\cal P}_Q$ plays an essential role for 
describing Mott physics.\cite{Miyagawa} 
Details will be given in the next subsection. 
\par

To accurately compute variational expectation values of $\Psi$, we use 
a variational Monte Carlo method in which the variational parameters are 
efficiently optimized using the quasi-Newton algorithm.\cite{qN} 
In some cases, we adopt the stochastic reconfiguration method\cite{SR} 
to reduce statistical fluctuation. 
We use $L\times L$ systems ($L=12$-$16$) with the periodic-antiperiodic 
boundary conditions. 
Although we implement calculations using the square lattice, the properties 
obtained for the normal state are qualitatively independent of the form and 
dimensionality of the lattice. 
In most cases, samples as many as $2.4\times 10^5$ are used to reduce 
numerical errors, which are typically $\sim 10^{-4}t$ for the total energy. 
\par
 
\subsection{Trial wave functions for repulsive cases\label{sec:wf}}
%
In this subsection, we mention a conventional part of wave functions used 
for systems without a Peierls phase $A$ in previous 
studies.\cite{YOT,doped_Mott}
For systems with $A$, we additionally need a phase factor ${\cal P}_\theta$, 
as we will discuss in Sec.~\ref{sec:P-theta}.
We study Mott transitions and Mott physics in a correlated $d$-wave singlet 
pairing state ($\Psi_d={\cal P}\Phi_d$) and a projected Fermi sea 
($\Psi_{\rm N}={\cal P}\Phi_\mathrm{F}$) as a normal state, and a crossover 
between a band insulator and a Mott insulator for a projected AF wave 
function ($\Psi_{\rm AF}={\cal P}\Phi_\mathrm{AF}$). 
\par

First, we explain the one-body part $\Phi$.
The BCS function with $d_{x^2-y^2}$- (uniform $s$-) wave gap is 
written as, 
%
\begin{eqnarray}
\Phi_{d(s)}=\left(\sum_{\bold{k}}a_\bold{k}c_{\bold{k}\uparrow}^\dag
c_{-\bold{k}\downarrow}^\dag\right)^{N_\mathrm{e}/2}|\mathrm{vac}
\rangle,\\
a_\bold{k}=\frac{\Delta_{d(s)}(\bold{k})}{\varepsilon_\bold{k}-\zeta
+\sqrt{(\varepsilon_\bold{k}-\zeta)^2+|\Delta_{d(s)}(\bold{k})|^2}}, 
\label{eq:BCS_d_wavefunc}
\end{eqnarray}
%
where $\varepsilon_\bold{k}=-2t(\cos k_x+\cos k_y)$, 
$\Delta_d(\bold{k})=\Delta(\cos k_x-\cos k_y)$ and $\Delta_s=\Delta$.
$\zeta$ and $\Delta$ are variational parameters which 
coincide with chemical potential and singlet pairing gap, respectively, 
in the limit of $U/t\rightarrow0$.
The Fermi sea (FS) is given by 
$\Phi_\mathrm{F}=\prod_{\bold{k}
\in \mathrm{FS},\sigma}c_{\bold{k}\sigma}^\dag|\mathrm{vac}\rangle$, 
and a mean-field-type AF state by
%
\begin{eqnarray}
\Phi_\mathrm{AF}=\prod_{\bold{k}\in\mathrm{FS},\sigma}\left[
\alpha_\bold{k}c_{\bold{k}\sigma}^\dag
+{\rm sgn}(\sigma)\beta_\bold{k}c_{\bold{k}+\bold{Q}\sigma}^\dag\right]
|\mathrm{vac}\rangle,\\
\alpha_\bold{k}(\beta_\bold{k})=\sqrt{\frac{1}{2}\left(
1-(+)\frac{\varepsilon_\bold{k}}{\sqrt{\varepsilon_\bold{k}^2
+\Delta_\mathrm{AF}^2}}
\right)}, 
\label{eq:AF_wavefunc}
\end{eqnarray}
where $\Delta_\mathrm{AF}$ is a variational parameter corresponding to 
the mean-field AF gap, and ${\rm sgn}(\sigma)=\pm 1$ according to 
$\sigma=\uparrow$ or $\downarrow$. 
\par

Next, we explain the many-body part ${\cal P}={\cal P}_\mathrm{G}
{\cal P}_Q$.
The most fundamental onsite (Gutzwiller) projector,\cite{Gutz} 
\begin{equation}
{\cal P}_\mathrm{G}=\prod_j\left[
1-(1-g)n_{j\uparrow}n_{j\downarrow}
\right], 
\label{eq:P_G}
\end{equation}
controls the density of doublons, 
$d\equiv\langle{\cal H}_{\rm int}\rangle/(UN_{\rm s}$); 
as the parameter $g$ decreases, $d$ decreases. 
Correspondingly, the range of $g$ is $0\leq g\leq 1$ 
($1\leq g \leq \infty$) for $U/t>0$ ($U/t<0$).
For $g=0$, doublons are completely excluded. 
For Mott physics, another correlation factor ${\cal P}_Q$ is crucial, 
which controls the binding between a doublon and a holon\cite{Kaplan,YS3} 
and is written explicitly as, 
\begin{eqnarray}
{\cal P}_Q&=&\prod_j(1-Q_j),\\
Q_j&=&\mu_dd_j\prod_\tau(1-h_{j+\tau})+\mu_hh_j\prod_\tau(1-d_{j+\tau}), 
\label{eq:P_Q}
\end{eqnarray}
where $d_j=n_{j\uparrow}n_{j\downarrow}$, 
$h_j=(1-n_{j\uparrow})(1-n_{j\downarrow})$ and $\tau$ runs nearest-neighbor 
sites of the site $j$. 
To treat doped (asymmetric) cases, we distinguish the contribution of a
D-to-H configuration from that of a H-to-D one,\cite{doped_Mott} which are 
controlled by the parameters $\mu_d$ and $\mu_h$, respectively. 
It was repeatedly shown that this type of short-range D-H binding factors 
capture the essence of Mott transition and Mott physics,\cite{Miyagawa} 
except for the Drude weight.\cite{M-C}
Although the form of Eq.~(\ref{eq:P_Q}) is slightly different from what 
was used by Millis and Coppersmith,\cite{M-C} the properties of the two 
are essentially the same. 
\par

\subsection{Drude weight and phase factor ${\cal P}_\theta$
\label{sec:P-theta}} 
%
According to Kohn,\cite{Kohn} in calculating the Drude weight, we add a 
vector potential $\bold{A}$ as a Peierls phase to ${\cal H}_{\rm kin}$ 
in Eq.(\ref{eq:H}): 
\begin{equation}
{\cal H}(\bold{A})=-t\sum_{<i,j>,\sigma}\left(
e^{-i\bold{A}\cdot(\bold{r}_i-\bold{r}_j)}c_{i\sigma}^\dag c_{j\sigma}
+\mathrm{H.c.} \right)+{\cal H}_{\rm int}.
\label{eq:H(A)}
\end{equation}
Assuming $\bold{A}=A\bold{\hat{x}}$, the Drude weight in $x$ direction is 
given by, 
\begin{equation}
D=\left.\frac{d^2E(A)}{dA^2}\right|_{A=0}, 
\label{eq:D=d^2E/dA^2}
\end{equation}
where $E(A)=\langle A|{\cal H}(A)|A\rangle$ with $|A\rangle$ being the 
normalized ground state of ${\cal H}(A)$.
Actually in this study, we obtain $D$ by calculating $[E(A)-E(0)]/2A^2$ 
for small $A$'s with $A\lesssim\pi/L$ ($L$: linear dimension of the system).
If the system is insulating [metallic], the relation $E(A)=E(0)$ 
[$E(A)\propto A^2$] holds, and $D=0$ [$D>0$].
\par

In constructing a trial state for ${\cal H}(A)$, we should notice that 
the matrix elements of Eq.~(\ref{eq:H(A)}) with respect to real-space 
configurations are complex. 
Because, in general, eigenvectors of an Hermitian matrix are essentially 
complex, a trial wave function should be complex.
To recognize the importance of imaginary part in the wave function, we apply 
Hellmann-Feynman's theorem in the second order to Eq.(\ref{eq:D=d^2E/dA^2}), 
\begin{eqnarray}
D&=&\langle 0|\left.\frac{d^2{\cal H}(A)}{dA^2}\right|_{A=0}|0\rangle+
2\langle 0|\left.\frac{d{\cal H}(A)}{dA}\right|_{A=0}
\left.\frac{d}{dA}|A\rangle\right|_{A=0} \nonumber\\
&=&\langle 0|t\sum_{i,\sigma}\left(c_{i\sigma}^\dag c_{i+\hat{x}\sigma}
+\mathrm{H.c.}\right)|0\rangle \nonumber\\
&&\:\:-2i\langle0|t\sum_{i,\sigma}(c_{i\sigma}^\dag 
                                                c_{i+\hat{x}\sigma}-\mathrm{H.c.})
\left.\frac{d}{dA}|A\rangle\right|_{A=0},
\label{eq:Hellmann-Feynman}
\end{eqnarray}
where $|0\rangle$ is the normalized ground state of ${\cal H}(0)$. 
Here we assume that the ground state of ${\cal H}(0)$ does not have 
a paramagnetic current, namely, $d\langle A|{\cal H}(A)|A\rangle/dA|_{A=0}=0$.
The first term of Eq.~(\ref{eq:Hellmann-Feynman}) is the absolute value of 
kinetic energy in $x$ direction, which is independent of $A$.
On the other hand, the second term depends on $A$ and includes imaginary 
unit $i$. 
When the system becomes insulating ($D=0$) with a finite kinetic energy, 
the first term must be cancelled out by the second term. 
Because $D$ is real, $|A\rangle$ must have a finite imaginary part. 
Thus, it is natural that the ordinary wave functions (${\cal P}\Phi$) 
discussed in Sec.~\ref{sec:wf}, including what was used by Millis and 
Coppersmith, exhibit metallic behavior for the Drude weight ($D>0$) for 
$U/t<\infty$, because these functions are real. 
\par

\begin{figure}[htbp]
\begin{center}
\includegraphics[width=3.5cm,clip]{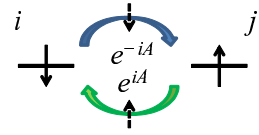} 
\end{center}
\vspace{-0.5cm} 
\caption{(Color online)
Schematic figure of an exchange process in Heisenberg model and attached 
phase factors. 
}
\label{fig:Heis}
\end{figure}
%
Now, let us construct a trial wave function suitable for a Mott insulating 
state with a finite $A$. 
We start with considering the AF Heisenberg model ($J>0$), 
\begin{equation}
{\cal H}_{\rm Heis}
=J\sum_{\langle i,j\rangle}\left[S_i^zS_j^z
+\frac{1}{2}\left(S^+_iS_j^-+S^-_iS_j^+\right)\right], 
\label{eq:Heis}
\end{equation}
which is the effective model in a Mott insulating regime ($U>U_{\rm c}$, 
$\delta=0$) of the Hubbard model. 
If a finite $A$ is applied---in this case the original Hamiltonian becomes 
Eq.~(\ref{eq:H(A)})---, the second-order virtual hopping processes 
in the strong-coupling expansion illustrated in Fig.~\ref{fig:Heis} does not 
yield a phase, because the two hopping processes occur sequentially, and 
phase factors cancel out each other: 
\begin{eqnarray}
e^{iA}c_{i\uparrow}^\dag c_{j\uparrow}
e^{-iA}c_{j\downarrow}^\dag c_{i\downarrow}
=c_{i\uparrow}^\dag c_{j\uparrow}c_{j\downarrow}^\dag c_{i\downarrow}. 
\end{eqnarray}
As a result, Eq.~(\ref{eq:Heis}) is invariant irrespective of whether $A$ 
is null or finite, and the matrix elements remain real. 
Thus, a phase factor is needless in a wave function for the Heisenberg 
model.\cite{Baeriswyl}
\par

\begin{figure}[htbp]
\begin{center}
\includegraphics[width=8.0cm,clip]{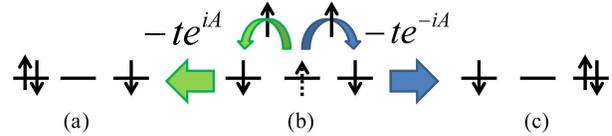} 
\end{center}
\caption{(Color online)
Illustration of hopping in ${\cal H}(A)$ in Mott-insulating region. 
A phase $e^{iA}$ [$e^{-iA}$] is added when an ($\uparrow$-spin) electron 
hops to the left (a) [right (c)] nearest-neighbor site from the original 
configuration (b).
}
\label{fig:pic_phase}
\end{figure}

On the other hand, in the corresponding virtual processes in the Hubbard 
model Eq.~(\ref{eq:H(A)}) with a large $U/t$---precisely, the two hopping 
processes in which a doublon is created and annihilated---, the two 
hoppings do not necessarily occur sequentially. 
Namely, we have to consider the two hoppings are mutually independent, 
although the two events occur in a paired manner (if not, the phase becomes 
metallic). 
Thus, we have to eliminate a phase added in a hopping process shown in 
Fig.~\ref{fig:pic_phase} within each process. 
To this end, we introduce into the wave function a configuration-dependent 
phase factor, 
\begin{equation}
{\cal P}_\theta=\exp\left[i\theta\sum_jd_j(h_{j+1}-h_{j-1})\right], 
\label{eq:Ptheta}
\end{equation}
where $j$ indicates the coordinate in $x$ direction, and $\theta$ is a 
variational parameter. 
If the relation $\theta=A$ is satisfied, the total phase factor vanishes 
in each hopping process. 
\par

This is also viewed from energetics in Eq.~(\ref{eq:D=d^2E/dA^2}). 
Assuming the trial wave function is real, kinetic energy increases 
proportionally to $-{\rm Re}(te^{\pm iA})$ when an ($\uparrow$-spin) 
electron in Fig.~\ref{fig:pic_phase}(b) hops to the right or left site. 
Because an insulating state satisfies $E(A)=E(0)$, this phase has to be 
cancelled by another phase factor such as Eq.~(\ref{eq:Ptheta}).
\par

If one applies $\Psi={\cal P}_\theta{\cal P}_Q{\cal P}_{\rm G}\Phi$ to 
${\cal H}(A)$ with $A=0$, the phase parameter $\theta$ in ${\cal P}_\theta$ 
is optimized at $\theta=0$, and ${\cal P}_\theta{\cal P}_Q{\cal P}_{\rm G}$ 
is reduced to ${\cal P}_Q{\cal P}_{\rm G}$ discussed in Sec.~\ref{sec:wf}.  
Thus, the results of quantities other than $D$ obtained in the previous 
studies are not modified by introducing ${\cal P}_\theta$. 
\par

Finally, we discuss some points related to ${\cal P}_\theta$.  
(1) The form of ${\cal P}_\theta$ in Eq.~(\ref{eq:Ptheta}) is the most 
fundamental one. 
It is possible to construct an improved trial state that takes account of 
wider electron configurations. 
We will return to this point in Sec.~\ref{sec:improved}. 
(2) ${\cal P}_\theta$, which is configuration dependent, is conceptually 
different from position-dependent phase factors used in different 
contexts.\cite{Liang,Weber,disorder} 
On the other hand, Millis and Coppersmith discussed the importance of 
configuration-dependent phase factors for insulating behavior in their 
original paper,\cite{M-C} but they did not provide a tractable form of 
wave functions. 
(3) In general, this type of configuration-dependent phase factor seems 
crucial for constructing current-carrying states in regimes of Mott 
physics in Hubbard-type models which allow double occupation. 
We return to this point in Sec.~\ref{sec:summary}. 
\par

\subsection{Drude and SC weights in variation 
theory\label{sec:D-and-Ds}}

The expression of the Drude weight $D$ in Eq.~(\ref{eq:D=d^2E/dA^2}) is 
identical to that of SC (Meissner) weight $D_{\rm s}$, but $D$ and 
$D_{\rm s}$ are different entities, in general. 
Scalapino, White and Zhang (SWZ)\cite{SWZ} proposed a criterion of 
distinguishing them in using Eq.~(\ref{eq:D=d^2E/dA^2}) by considering 
the possibility of level crossings at small $A$'s: $D$ is given by the 
$A$ derivative of the ground-state energy at $A=0$ (adiabatic derivative), 
while $D_{\rm s}$ is given by the $A\rightarrow 0$ limiting value of $A$ 
derivative of the ground-state energy at finite $A$'s (envelope derivative). 
\par

However, this criterion has subtle points,\cite{Hetenyi} and is not directly 
applicable to variation theory, because, generally, a trial function is not 
assumed to undergo frequent level crossings for $A\sim 0$. 
If one sets a normal (SC) state as a trial state at $A=0$, the state 
remains normal (SC) at a finite $A$ in most cases. 
Therefore, henceforth, we simply regard the quantity obtained using 
Eq.~(\ref{eq:D=d^2E/dA^2}) as $D$ ($D_{\rm s}$), if a trial state is 
incoherent (coherent) such as $\Psi_{\rm N}$ and $\Psi_{\rm AF}$ ($\Psi_d$ 
and $\Psi_s$). 
We consider $D=D_{\rm s}$ for coherent states, and $D_{\rm s}=0$ for 
incoherent states, following common relations.\cite{SWZ} 
\par

\section{$d$-wave pairing and normal states for $U>0$\label{sec:normalSC}}
In Sec.~\ref{sec:plain}, we discuss the results of $d$-wave pairing state 
($\Psi_d$) and normal state ($\Psi_\mathrm{N}$) with 
${\cal P}={\cal P}_\theta{\cal P}_Q{\cal P}_\mathrm{G}$. 
It is known that $\Psi_d$ ($\Psi_\mathrm{N}$) exhibits a first-order Mott 
transition at $U_c/t\simeq 6.5$ ($8.6$)\cite{YTOT,YOT} in quantities 
such as $d$, $n({\bf k})$ and $N({\bf q})$. 
We mainly display the results of $\Psi_d$, because the behavior of $D$ and 
Mott transitions is qualitatively similar between $\Psi_d$ and 
$\Psi_\mathrm{N}$. 
In Sec.~\ref{sec:improved}, we improve the phase factor upon 
${\cal P}_\theta$. 
\par

\subsection{Plain phase factor ${\cal P}_\theta$\label{sec:plain}}
\begin{figure*}[htbp]
\begin{center}
\includegraphics[width=14cm,clip]{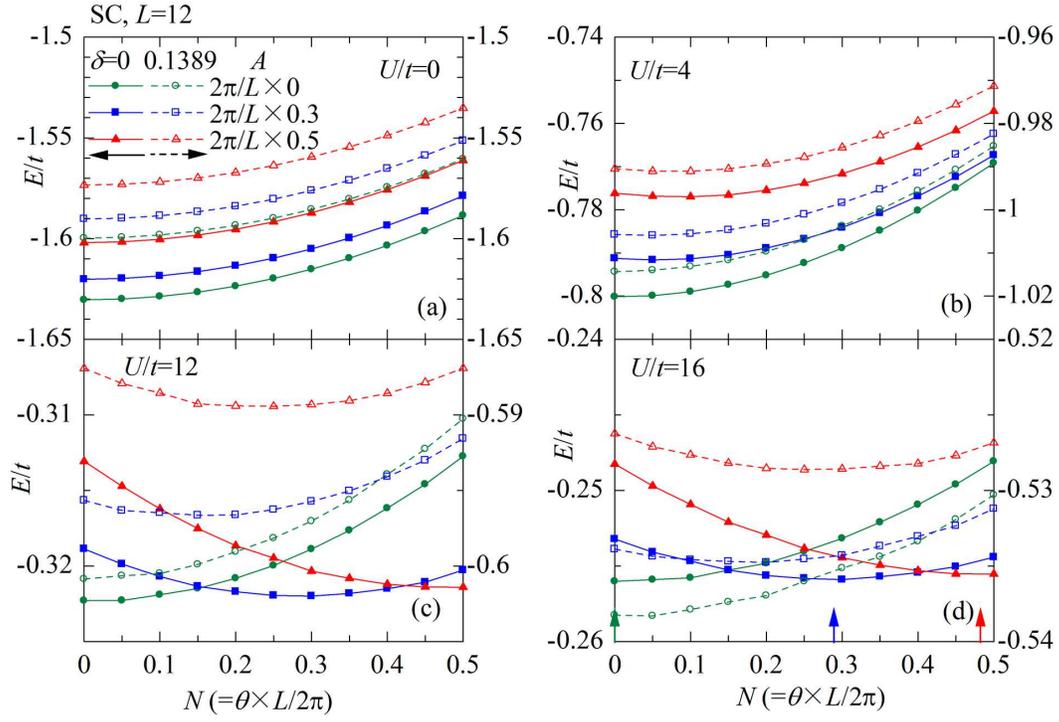} 
\end{center}
\caption{(Color online)
Total energy per site is plotted as a function of the phase parameter 
$\theta$ for the $d$-wave pairing wave function $\Psi_d$. 
In each panel, data for a couple of doping rates and vector potentials are 
shown; among the four panels, the value of $U/t$ is different. 
We scale $\theta$ to $N=\theta\times L/2\pi=\theta\times 12/2\pi$ in abscissa.
The arrows in (d) indicate the optimized values of $\theta$ 
($\theta_{\rm min}$) for $\delta=0$ and the three values of $A$.
}
\vspace{-0.3cm} 
\label{fig:Energy_phi}
\end{figure*}
%
\begin{figure}[htbp]
\begin{center}
\includegraphics[width=8.0cm,clip]{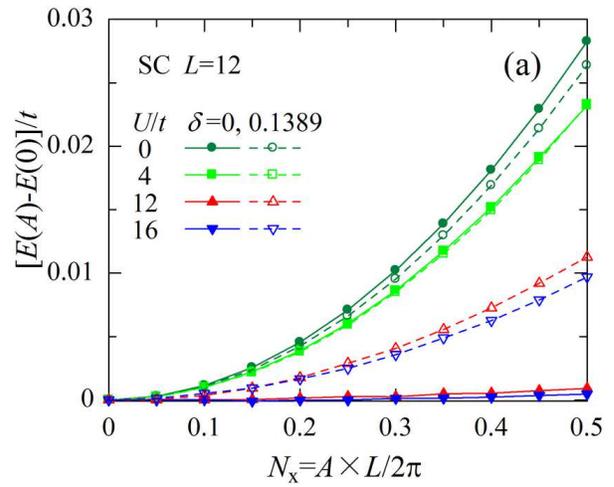} 
\end{center}
\vspace{-0.5cm} 
\caption{(Color online)
Increment in total energy when a vector potential $A$ is applied is shown 
as a function of $A$. 
Cases of four values of $U/t$ are compared for $\delta=0$ (half filling) 
and $\delta=0.1389$. 
$U_{\rm c}/t\sim 6.5$.
}
\vspace{-0.3cm} 
\label{fig:E_A}
\end{figure}
%
\begin{figure}[htbp]
\begin{center}
\includegraphics[width=8.5cm,clip]{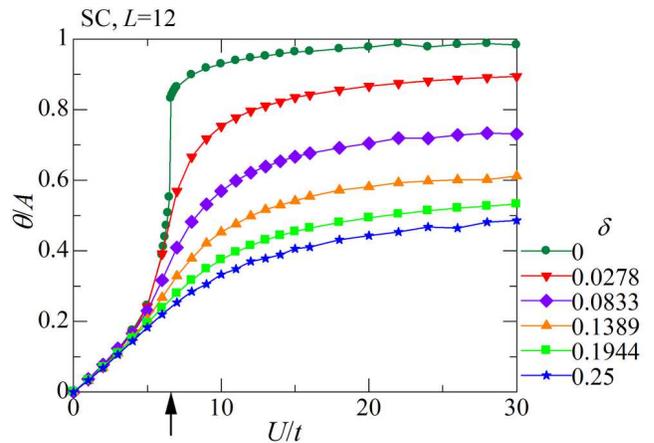} 
\end{center}
\vspace{-0.5cm} 
\caption{(Color online)
The optimized phase parameter $\theta$ is plotted as a function of $U/t$ 
for several doping rates. 
The arrow indicates the Mott transition point at half filling.
}
\label{fig:theta_A}
\end{figure}
%
First, we show how energy is reduced by introducing the phase parameter 
$\theta$. 
In Fig.~\ref{fig:Energy_phi}, we plot total energy per site $E$ for a few 
values of small $A$ and $\delta$. 
Here, all variational parameters other than $\theta$ are optimized. 
Shown in the panels (a) and (b) are the cases of weak correlations, 
$U/t=0$ and $4$, respectively, in which the systems are metallic 
($U<U_{\rm c}$), as mentioned. 
Let the optimized $\theta$ be $\theta_\mathrm{min}$. 
We find $\theta_\mathrm{min}$ is situated at or in the very vicinity of 
zero for any values of $A$ and $\delta$. 
As a result, $E(A)$ basically becomes an increasing function of $A$. 
On the other hand, in a strongly correlated regime ($U>U_{\rm c}$) 
[Figs.~\ref{fig:Energy_phi}(c) and (d)], $\theta_\mathrm{min}$ shifts 
to an appreciably large value as $A$ increases [see the arrows in 
Fig.~\ref{fig:Energy_phi}(d)]. 
However, the value of $E(A)$ at $\theta=\theta_\mathrm{min}$ changes only 
very slightly when $A$ is varied. 
This situation is summed up in Fig.~\ref{fig:E_A}, where we plot the 
energy increment when a small $A$ is applied. 
Here, all the variational parameters including $\theta$ are optimized. 
For $U/t=0$ and a large $L$, $E(A)-E(0)$ is expanded with respect to $A$ 
in a quadratic function as, 
\begin{equation}
E_0(A)-E_0(0)=A^2t\left(\frac{1}{2\pi}\right)^2
\int_{{\bf k}\in{\bf k}_{\rm F}}d{\bf k}\cos k_x+\cdots. 
\label{eq:increment}
\end{equation}
\par

\begin{figure*}[htbp]
\begin{center}
\includegraphics[width=15cm,clip]{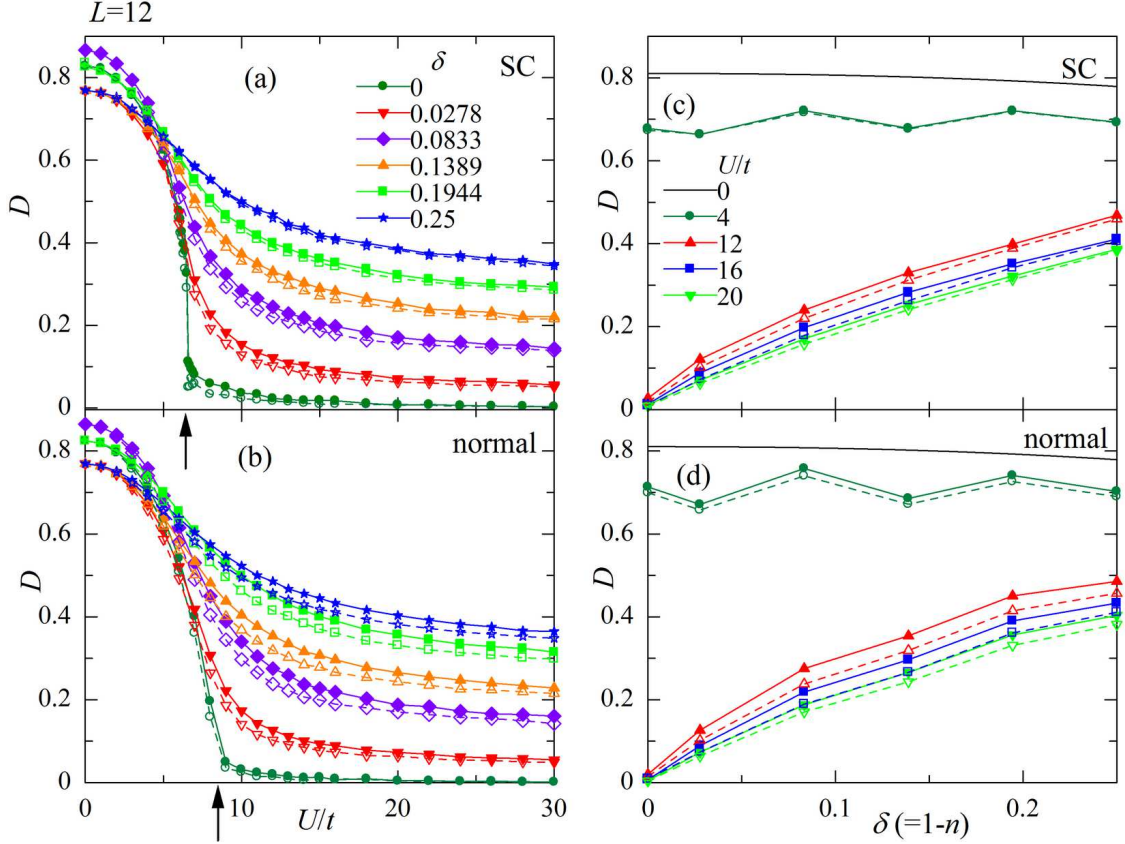} 
\end{center}
\vspace{-0.4cm} 
\caption{(Color online) 
In (a) and (b), the Drude weight is plotted as a function of $U/t$ for the 
$d$-wave pairing and normal states, respectively. 
The symbols are common. 
In each panel, data for several $\delta$'s are plotted. 
The arrows indicate the Mott transition points at half filling. 
For the normal state, $U_{\rm c}/t\sim 8.7$.\cite{YOT}
In (c) and (d), the same quantity is shown as a function of doping rate 
for the same states, respectively. 
In each panel, finite-$U/t$ VMC data ($L=12$) are shown with symbols; 
Solid symbols and solid lines denote the results of ${\cal P}_\theta\Psi$ 
and open symbols and dashed lines those of $\tilde{\cal P}_\theta\Psi$ 
discussed in Sec.~\ref{sec:improved}. 
For comparison, the behavior of $D$ for $U/t=0$ ($L=\infty$) is plotted 
with a black solid line. 
}
\vspace{-0.2cm} 
\label{fig:Drude_dope}
\end{figure*}

\begin{figure}[htbp]
\begin{center}
\includegraphics[width=8.5cm,clip]{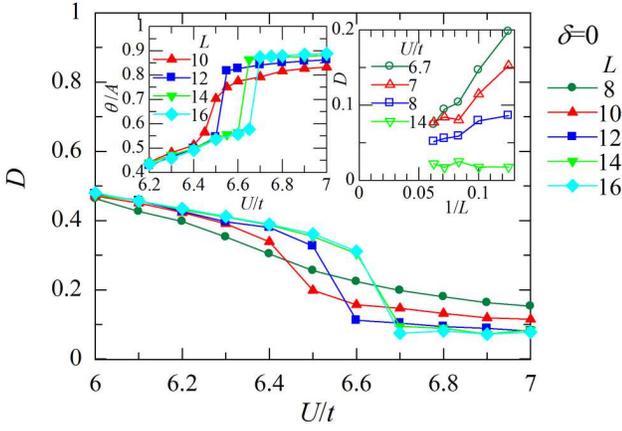} 
\end{center}
\vspace{-0.5cm} 
\caption{(Color online) 
Magnification of Drude weight for $\delta=0$ in $d$-wave pairing state 
near Mott transition points $U_{\rm c}/t\sim 6.5$ for several values of $L$.
In the right inset, the system-size dependence of $D$ is shown for four 
values of $U/t$ in the Mott-insulating regime. 
We estimate $D$ with $A=2\pi/L\times0.2$. 
In the left inset, the system-size dependence of optimized $\theta$ is 
plotted versus $U/t$. 
}
\vspace{-0.3cm} 
\label{fig:D_size}
\end{figure}

First, we consider the half-filled case (solid symbols in Fig.~\ref{fig:E_A}). 
We find this quadratic behavior still continues in the metallic regime 
($U/t=4$). 
On the other hand, in the insulating regime ($U/t=12$, $16$), $E(A)$ becomes 
almost constant [$=E(0)$], as mentioned. 
Thus, the behavior of $E(A)-E(0)$ qualitatively changes through $U_{\rm c}/t$. 
To pursue it, we show, in Fig.~\ref{fig:theta_A}, the optimized phase 
parameter $\theta$ as a function of $U/t$. 
For $\delta=0$, $\theta$ exhibits a discontinuity at $U=U_{\rm c}\sim 6.5t$. 
We confirmed that this value of $U_{\rm c}/t$ precisely coincides with 
that previously estimated from other quantities.\cite{YTOT,YOT} 
Note that, for $U>U_{\rm c}$, $\theta$ approaches $A$, and increases as $L$ 
increases as shown in the left inset of Fig.~\ref{fig:D_size}, in contrast 
to the behavior for $U<U_{\rm c}$, where $\theta$ decreases with $L$. 
It demonstrates that the mechanism assumed when the phase factor 
${\cal P}_\theta$ was introduced [Eq.~(\ref{eq:Ptheta})] certainly works 
for $U\gtrsim U_{\rm c}$. 
\par 

We turn to doped cases.  
As shown in Fig.~\ref{fig:E_A}, $E(A)-E(0)$ continues to be a quadratic 
function of $A$ even for sufficiently large values of $U/t$. 
In Fig.~\ref{fig:theta_A}, no discontinuity in the behavior of optimized 
$\theta$ is found even for the smallest doped case ($\delta=0.0278$). 
Thus, the Mott transition vanishes on doping, as expected. 
As $\delta$ increases, the optimized value of $\theta$ decreases from $A$. 
This is mainly because an isolated (untied to doublon) doped hole adds a 
phase $A$ or $-A$ during the hopping, but ${\cal P}_\theta$ does not 
compensate for it. 
In general, it is probably impossible to make a phase factor which 
compensates for the phase generated by a free carrier. 
We may say this is the origin of conduction. 
Another reason for $\theta<A$ is interplay of a doublon and multiple 
holons; we will take up this effect in Sec.~\ref{sec:improved}. 
\par

%
%
Using $E(A)$ obtained above, we estimate the Drude weight through 
Eq.~(\ref{eq:D=d^2E/dA^2}). 
In Fig.~\ref{fig:Drude_dope}, we plot $D$ for the $d$-wave pairing and 
normal states (solid symbols). 
To begin with, we consider $U/t$ dependence [panels (a) and (b)] at half 
filling. 
For $U/t=0$, as derived from Eq.~(\ref{eq:increment}), the Drude weight 
becomes, 
\begin{equation}
D=\frac{|E_{\rm kin}^x|}{t}, 
\label{eq:D-U0}
\end{equation}
where $E_{\rm kin}^x$ is the kinetic energy in $x$ direction. 
Thus, we have $D=8/\pi^2$ at $\delta=0$ for the square lattice. 
As $U/t$ increases, $D$ rapidly decreases and almost vanishes at 
$U=U_{\rm c}$ in both states as shown by arrows in 
Figs.~\ref{fig:Drude_dope}(a) and \ref{fig:Drude_dope}(b). 
In Fig.~\ref{fig:D_size}, the behavior of $D$ near the Mott transition point 
is compared among different $L$'s. 
It reveals $D$ exhibits a discontinuity at $U=U_{\rm c}$ for large 
$L$'s ($L\gtrsim 12$).\cite{note-size}
Note that $D$ decreases as $L$ increases for $U>U_{\rm c}$, in contrast 
to the feature for $U<U_{\rm c}$. 
In the right inset of Fig.~\ref{fig:D_size}, $D$ is plotted as a function of 
$1/L$ for some values of $U/t$ for $U>U_{\rm c}$. 
Although accurate extrapolation is not easy owing to the statistical 
fluctuation for large $L$'s, $D$ probably vanishes for $L\rightarrow\infty$. 
The value of $U_{\rm c}/t$ and the features of a first-order transition 
precisely coincide with those obtained in other quantities.\cite{YOT} 
Thus, we could show for $\Psi_d$ and $\Psi_{\rm N}$ that a conductor and 
a Mott insulator can be clearly distinguished by the Drude weight, using a 
variation theory.  
\par
\begin{figure}[htbp]
\begin{center}
\vspace{0.5cm} 
\includegraphics[width=7.5cm,clip]{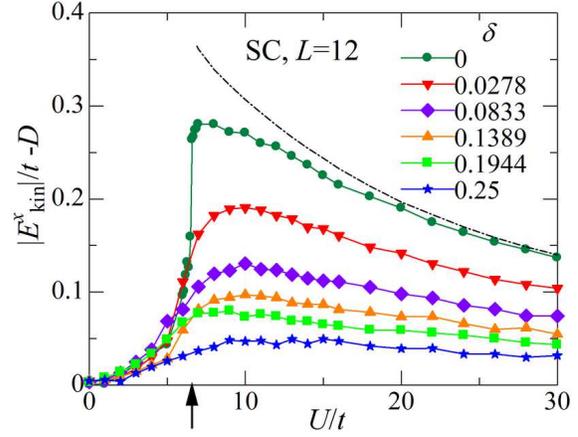} 
\end{center}
\vspace{-0.5cm} 
\caption{(Color online) 
Contribution of the second term in Eq.~(\ref{eq:Hellmann-Feynman}), namely, 
the difference between the absolute value of kinetic energy in $x$ direction 
and $D$ is plotted for $\Psi_d$ as a function of $U/t$ for several doping 
rates. 
For comparison, $|E_{\rm kin}^x|/t$ for $\delta=0$ is shown with a 
dash-dotted line. 
The arrow indicates the Mott transition point at half filling. 
}
\label{fig:Ex-D}
\end{figure} 
%
Now, we look at the effectiveness of ${\cal P}_\theta$ with respect to $U/t$ 
and $\delta$. 
As discussed, without the imaginary part in $\Psi$, $D$ is reduced to 
$|E_{\rm kin}^x|/t$ [the first term in Eq.~(\ref{eq:Hellmann-Feynman})], 
which is proportional to $t/U$ for $U/t\rightarrow\infty$. 
The contribution of the second term in Eq.~(\ref{eq:Hellmann-Feynman}), 
which is the direct effect of ${\cal P}_\theta$, is given by 
$|E_{\rm kin}^x|/t-D$. 
In Fig.~\ref{fig:Ex-D}, $U/t$ dependence of this quantity is shown for some 
$\delta$'s. 
At half filling, this quantity is very small compared with 
$|E_{\rm kin}^x|/t$ for $U<U_{\rm c}$, whereas the two quantities approach 
each other for $U>U_{\rm c}$. 
As $\delta$ increases, the contribution of $|E_{\rm kin}^x|/t-D$ for 
$U>U_{\rm c}$ rapidly becomes weak, in contrast to the increase in 
$|E_{\rm kin}^x|/t$. 
Consequently, $D$ increases as $\delta$ increases, as we see next. 
\par

Shown in Figs.~\ref{fig:Drude_dope}(c) and \ref{fig:Drude_dope}(d) is 
the doping dependence of $D$. 
For $U/t=0$ (black solid line), the Drude weight starts from $8/\pi^2$ 
at $\delta=0$ and monotonically and slowly decreases as $\delta$ increases, 
because Eq.~(\ref{eq:D-U0}) holds. 
For $U<U_{\rm c}$, this weak dependence on $\delta$ continues as seen for 
$U/t=4$. 
Here, the alternate behavior for a small $U/t$ stems from the boundary 
conditions we use (a finite-size effect), and is irrelevant. 
The behavior of $D$ abruptly changes around $U=U_{\rm c}$. 
For $U>U_{\rm c}$, $D$ increases linearly as $\delta$ increases. 
Difference is slight between the normal and $d$-wave SC states.
\par

Linear behavior of $D(\delta)$ was observed for the $t$-$J$ 
model,\cite{Paramekanti} and therefore this behavior is characteristic of 
doped Mott insulators. 
Also in strongly correlated Hubbard models, the behavior of $D(\delta)$, 
namely, $D\propto \delta^\alpha$ for $\delta\rightarrow 0$ has been studied 
for long, mainly on the basis of exact 
diagonalization.\cite{Dagotto,Tohyama,Tohyama2,Nakano}
The main concern of these studies was whether the exponent $\alpha$ is 
1\cite{Tohyama,Tohyama2} or 2.\cite{Nakano} 
On the other hand, it is well-known that linear behavior of $D(\delta)$ 
was found in the cuprate superconductors first by $\mu$SR experiments 
(so-called Uemura plot),\cite{Uemura} which gave strong evidence that 
the cuprate SC's are doped Mott insulators. 
A recent experiment\cite{Tallon} showed SC weight $\rho_{\rm s}$ looks 
somewhat convex, which is similar to $D_{\rm s}$ for $U>U_{\rm c}$ in 
Fig.~\ref{fig:Drude_dope}(c). 
Although the relationship of the present results to experiments has to be 
deliberately analyzed, we believe both capture a typical feature of 
Mott physics.
\par

\subsection{Improved phase factor\label{sec:improved}}
%
\begin{figure}[htbp]
\begin{center}
\includegraphics[width=8.5cm,clip]{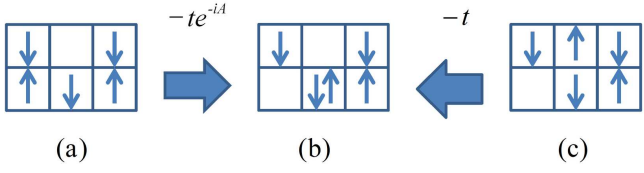} 
\vspace{-0.3cm} 
\end{center}
\caption{(Color online) 
Difference of attached phases is illustrated between hopping processes 
from (a) to (b) and from (c) to (b), which make a doublon with two 
neighboring holons in $x$ and $y$ directions [(b)]. 
}
\label{fig:n_lower}
\end{figure}
%
Average distance between a doublon and a holon becomes appreciably larger 
than the nearest-neighbor-site distance for 
$U\lesssim U_{\rm c}$.\cite{Miyagawa} 
However, when we construct ${\cal P}_\theta$, we only take account of 
the contribution of nearest-neighbor D-H pairs. 
In addition, when holes are doped, the interplay of a doublon and multiple 
holons becomes effective. 
As shown in Fig. \ref{fig:n_lower}(b), a doublon is frequently next to two 
holons in $x$ and $y$ directions simultaneously for $\delta>0$. 
This configuration is realized by a single hopping from a configuration 
in Fig.~\ref{fig:n_lower}(a) or Fig.~\ref{fig:n_lower}(c). 
Although these two hoppings occur with the same possibility if $A$ is 
sufficiently small, added phases are different between the two, namely, 
$e^{-iA}$ ($e^0$) is added in the former (latter) hopping. 
On the other hand, the phase factor ${\cal P}_\theta$ attaches the same 
counter-phase $\theta$ to the two hopping processes. 
Thus, even without considering the effect of isolated holons, 
$\theta_{\rm min}$ deviates from $A$ as $U/t$ becomes small or $\delta$ 
increases. 
It is better to treat a D-H pair in $x$ direction independently of a pair 
in $y$-direction. 
\par

\begin{figure}[htbp]
\begin{center}
\includegraphics[width=7cm,clip]{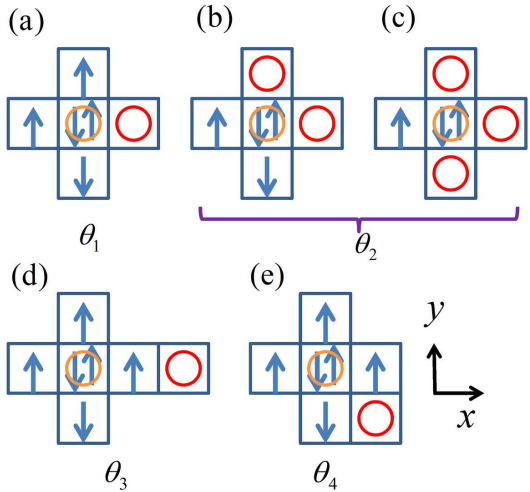} 
\end{center}
\vspace{-0.5cm} 
\caption{(Color online) 
We illustrate the way of assigning phase factors $e^{i\theta_j}$ 
$(j=1,2,\cdots,8)$ to local electron configurations in the improved phase 
factor, $\tilde{\cal P}_\theta$. 
Details are explained in the text. 
}
\label{fig:P_long_pic}
\end{figure}
%
On the basis of this argument, we substitute an improved projection factor 
$\tilde{\cal P}_\theta$ for ${\cal P}_\theta$. 
In $\tilde{\cal P}_\theta$, we allow for multiple-holon effects as well as 
extended D-H binding effects up to the two-step distance with eight phase 
parameters ($\theta_1$-$\theta_8$) to be optimized. 
Because the operator representation of $\tilde{P}_\theta$ is complicated, 
we explain the function of $\tilde{P}_\theta$ with illustrations in 
Fig.~\ref{fig:P_long_pic}. 
Formally, $\tilde{P}_\theta$ is represented as
\begin{equation}
\tilde{\cal P}_\theta=\prod_\ell \left(e^{i\theta_j}\right)_\ell, 
\label{eq:tilde-P}
\end{equation}
where $\ell$ is a site index. 
For example, if a doublon at site $\ell$ (orange circle in 
Fig.~\ref{fig:P_long_pic}) has a nearest-neighbor holon as in 
Fig.~\ref{fig:P_long_pic}(a), $e^{i\theta_1}$ is assigned. 
If a doublon has no nearest-neighbor holon but a second-neighbor holon 
as in Fig.~\ref{fig:P_long_pic}(d), $e^{i\theta_3}$ is assigned. 
If a doublon has no first- and second-neighbor holon, as well as if site 
$\ell$ is empty or singly occupied, we assign $e^{i0}$. 
In Fig.~\ref{fig:P_long_pic}, we only explain D-to-H factors 
($\theta_1$-$\theta_4$), but we also consider H-to-D factors 
($\theta_5$-$\theta_8$) in the same way. 
In Fig.~\ref{fig:P_long_pic}, we only show configurations in which D is 
located on the left to H; when the positions of D and H are exchanged, 
an assigned phase factor should be $e^{-i\theta_j}$. 
\par

\begin{table}
\caption{
Improvement of total energy for the normal state ($L=12$) by 
$\tilde{{\cal P}}_\theta$ on ${\cal P}_\theta$ is summarized, when a Peierls 
phase $A$ [$=0.5 L/(2\pi)$] is applied for some values of $U/t$ and $\delta$. 
The second, third, and fourth rows indicate the total energies of the 
given wave functions. 
The values in brackets indicate errors in the last digits. 
The last low shows the ratio 
$[E(\Psi)-E({\cal P}_\theta\Psi)]/
[E(\Psi)-E(\tilde{{\cal P}}_\theta\Psi)]$.
}
\begin{tabular}{c|c|c|c|c}
\hline
$\ U/t,\ \delta\ $ & 
$4,\ 0.0$      & $12,\ 0.0$    & $12,\ 0.1389$ & $ 16,\ 0.0$   \\ 
\hline\hline
$\Psi_{\rm N}$ & 
$-0.77484(7)$ & $-0.2538(2)$ & $-0.5779(1)$   & $-0.2029(2)$ \\ 
\hline
${\cal P}_\theta\Psi_{\rm N}$ & 
$-0.77551(7)$ & $-0.2608(9)$ & $-0.5796(1)$   & $-0.2092(1)$ \\ 
\hline
$\tilde{\cal P}_\theta\Psi_{\rm N}$ & 
$-0.77599(7)$ & $-0.2610(2)$ & $-0.5808(1)$   & $-0.2091(2)$ \\
\hline\hline
\mbox{Ratio}  & 
$0.58$        & $0.97$       & $0.59$         & $1.00$       \\
\hline
\end{tabular} 
\label{tb:Drude_comp}
\end{table}
%
The Drude weight calculated with $\tilde{\cal P}_\theta\Psi$ is shown 
in Fig. \ref{fig:Drude_dope} with open symbols. 
As compared to the results of simple ${\cal P}_\theta\Psi$ (solid symbols), 
the magnitude of $D$ becomes small for any values of $U/t$ and $\delta$. 
For a quantitative analysis, we compare, in Table~\ref{tb:Drude_comp}, the 
total energy for a small $A$ among three wave functions, namely, 
$\Psi_{\rm N}$ without a phase factor, ${\cal P}_\theta\Psi_{\rm N}$ and 
$\tilde{\cal P}_\theta\Psi_{\rm N}$ for typical values of $U/t$ and $\delta$. 
In the last row, we show the ratio of improvement by ${\cal P}_\theta\Psi$ 
to the improvement by $\tilde{\cal P}_\theta\Psi$. 
Although the statistical fluctuation is large, we can grasp a tendency. 
The improvement by $\tilde{\cal P}_\theta\Psi$ is notable for an intermediate 
$U/t$ or a finite $\delta$; not a small portion [$\sim 40\%$ for 
($U/t$, $\delta$)=($4$, $0.0$) and ($12$, $0.1389$)] of the improvement is 
achieved by the newly introduced part in $\tilde{\cal P}_\theta\Psi$. 
On the other hand, for $\delta=0$ and $U>U_{\rm c}$, the difference between 
$E({\cal P}_\theta\Psi)$ and $E(\tilde{\cal P}_\theta\Psi)$ is negligible. 
This result is what we expected. 
\par

In summary, the essence of Mott transitions as to the Drude weight is 
captured by the simple phase factor ${\cal P}_\theta$, but quantitative 
improvement is possible for intermediate correlation strengths or finite 
doping rates by introducing refined phase factors such as 
$\tilde{\cal P}_\theta$. 
\par

\section{Antiferromagnetic state\label{sec:AF}}
%
In this section, we study the Drude weight of AF states. 
In contrast to the normal and $d$-wave pairing states, which are insulating 
only for $U>U_{\rm c}$ at half filling, the AF states are insulating for any 
positive value of $U/t$ for $\delta=0$, because the nesting condition is 
completely satisfied for hypercubic lattices (cf. Fig.~\ref{fig:nk-nq}). 
Thus, the nature of AF states as insulators changes from a band (Slater) 
insulator to a Mott insulator as $U/t$ increases. 
As we will see shortly, AF states become the touchstone of applicability 
of the phase factor ${\cal P}_\theta$. 
\par

\begin{figure}[htbp]
\begin{center}
\includegraphics[width=7.5cm,clip]{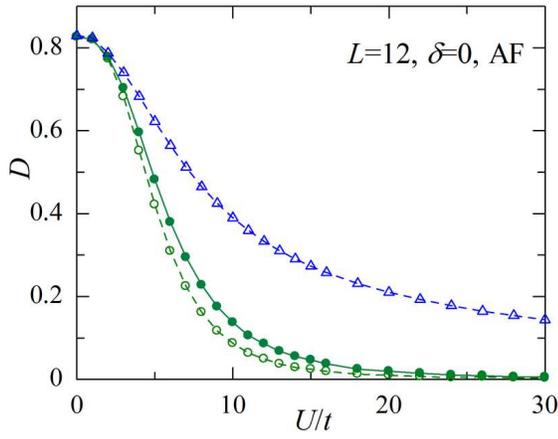} 
\end{center}
\vspace{-0.3cm} 
\caption{(Color online) 
The Drude weight of $\Psi_{\rm AF}$ at half filling is shown as a function 
of $U/t$.
Solid (open) circles indicate VMC results of using ${\cal P}_\theta$ 
($\tilde{\cal P}_\theta$) as a phase factor. 
For comparison, $|E_{\rm kin}^x|/t$ is shown with open triangles. 
}
\label{fig:Drude_AF}
\end{figure}
%
\begin{figure}[htbp]
\begin{center}
\includegraphics[width=7.5cm,clip]{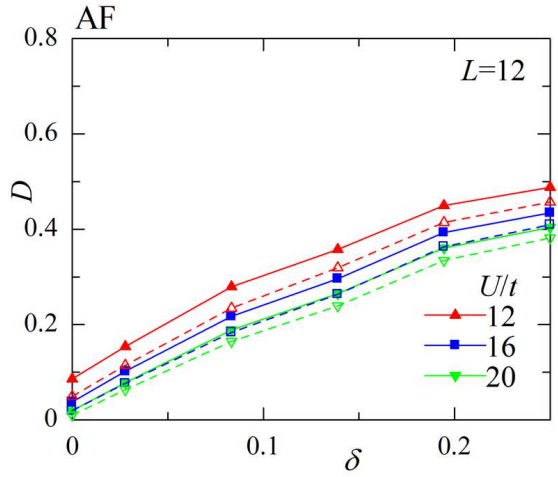} 
\end{center}
\vspace{-0.3cm} 
\caption{(Color online)
Drude weight of $\Psi_{\rm AF}$ calculated using VMC as a function of 
doping rate.
Solid (open) symbols indicate the results of using 
${\cal P}_\theta$ ($\tilde{\cal P}_\theta$) as a phase factor.
For large $\delta$'s (e.g.~$\delta\gtrsim 0.15$ for $U/t=12$), the state 
becomes normal. 
}
\label{fig:Drude_AF_dope}
\end{figure}
%
\begin{figure}[htbp]
\begin{center}
\includegraphics[width=7.0cm,clip]{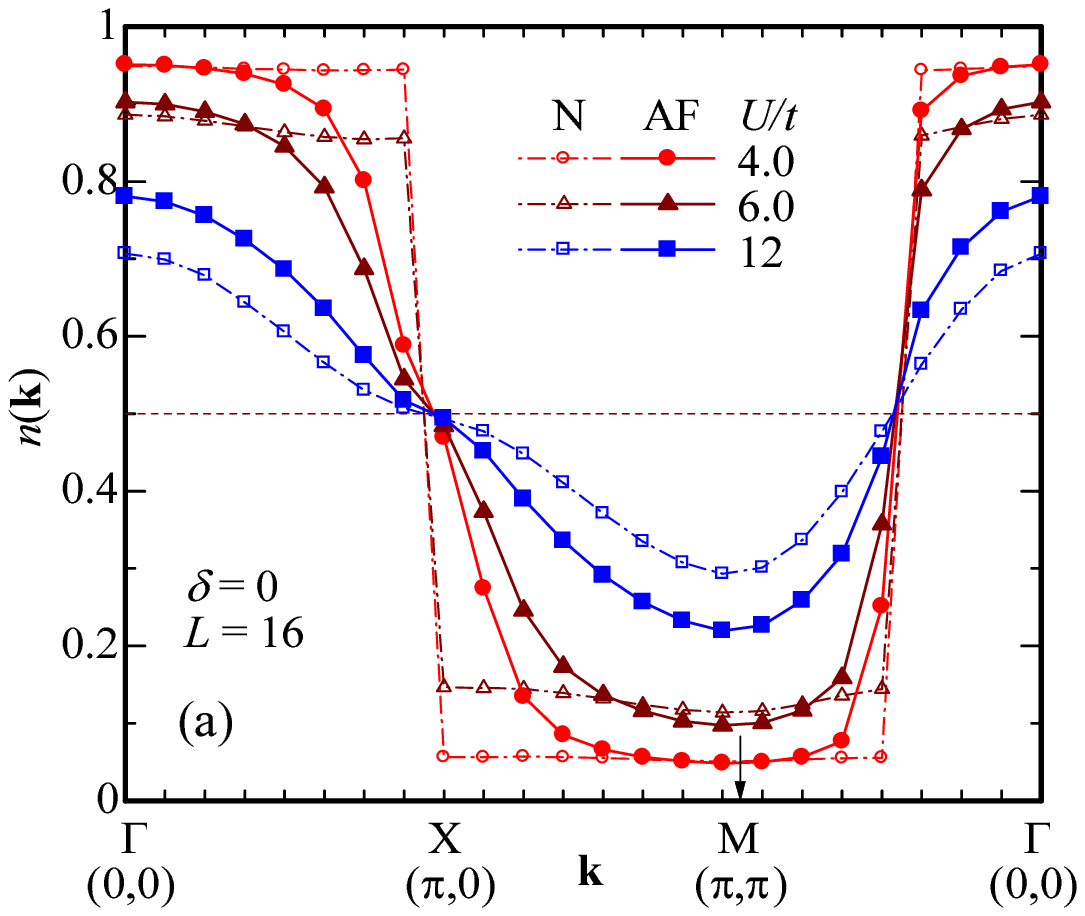} 
\includegraphics[width=7.0cm,clip]{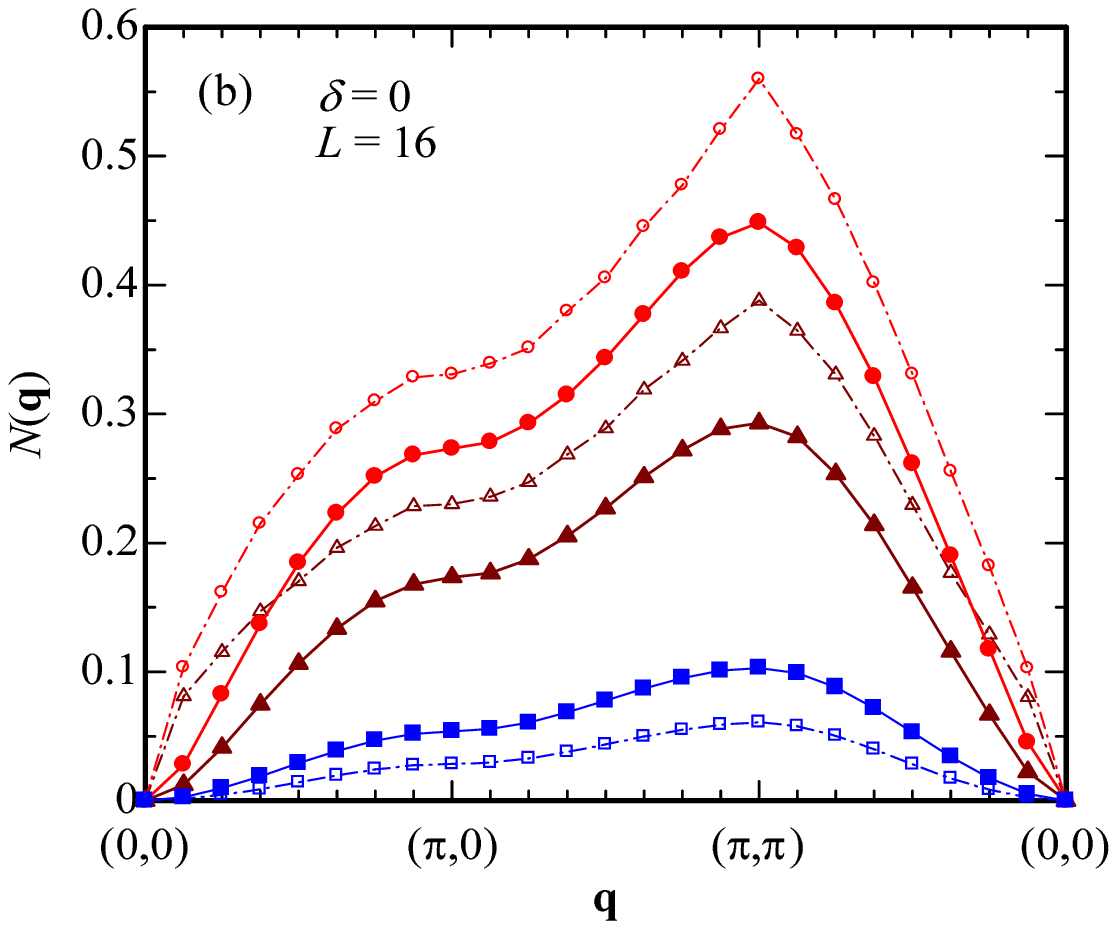} 
\end{center}
\vspace{-0.5cm} 
\caption{(Color online)
(a) Momentum distribution function and (b) charge density structure factor 
of $\Psi_{\rm AF}$ (solid symbols) and $\Psi_{\rm N}$ (open symbols) 
calculated using VMC are plotted along the path 
$(0,0)$-$(\pi,0)$-$(\pi,\pi)$-$(0,0)$ for three values of $U/t$. 
The Mott transition point for $\Psi_{\rm N}$ is $U_{\rm c}/t\sim 8.7$.
}
\label{fig:nk-nq}
\end{figure}
%
Let us start with the results of $\Psi_{\rm AF}$ obtained by VMC. 
In Fig.~\ref{fig:Drude_AF}, we plot $U/t$ dependence of $D$ at half filling. 
Similarly to $\Psi_{\rm N}$ and $\Psi_{d}$ [Figs.~\ref{fig:Drude_dope}(a) 
and \ref{fig:Drude_dope}(b)], $D$ starts from $D=8/\pi^2$ for $U/t=0$ and 
decreases, as $U/t$ increases, rapidly but smoothly owing to absence of 
a transition. 
First, we discuss the strongly correlated regime. 
For $U/t\gtrsim 10$, $D$ becomes very small, compared with 
$|E_{\rm kin}^x|/t$, indicating that ${\cal P}_\theta$ appropriately works 
also for $\Psi_{\rm AF}$ in this regime. 
Doping-rate dependence of the Drude weight shown in 
Fig.~\ref{fig:Drude_AF_dope} is almost linear, and similar to those for 
$\Psi_{\rm N}$ and $\Psi_d$ [Figs.~\ref{fig:Drude_dope}(c) 
and \ref{fig:Drude_dope}(d)]. 
Next, we consider the weakly correlated regime ($U\lesssim W$), where 
$\Psi_{\rm AF}$ is a band insulator. 
Being an insulator is confirmed by the behavior of the momentum distribution 
function and charge density structure factor shown in Fig.~\ref{fig:nk-nq}. 
Namely, there is no Fermi surface [no discontinuity in $n({\bf k})$], and 
a charge gap opens [$N({\bf q})\propto q^\alpha$ with $\alpha>1$ for 
$|{\bf q}|\rightarrow 0$].
Therefore, $D$ should vanish even for $U\lesssim W$. 
However, the Drude weight calculated with $\Psi_{\rm AF}$ exhibits large 
finite values for $U\lesssim W$. 
This indicates that $\Psi_{\rm AF}$ is lacking in some important element 
for describing $D$ in a band insulator. 
\par 

It is useful to consider this point with a mean-field theory, assuming 
$U/t$ is small. 
Note that the one-body part $\Phi_\mathrm{AF}$ is an insulating wave 
function for $U/t>0$ at half filling, even without correlation factor 
${\cal P}$. 
Therefore, $D$ has to vanish within the mean-field theory, meaning 
the one-body part has to be complex. 
Because $\Phi_\mathrm{AF}$ is essentially real, we need to modify it 
in this context. 
Let us look at this point in the light of a perturbation theory. 
We expand ${\cal H}(A)$ in Eq.~(\ref{eq:H(A)}) with respect to $A$ as, 
\begin{equation}
{\cal H}(A)\simeq{\cal H}(0)+AJ_x,
\label{eq:H-AJ}
\end{equation}
with paramagnetic current
\begin{equation}
J_x=-it\sum_{\bold{r},\sigma}(c_{\bold{r}\sigma}^\dag 
c_{\bold{r+\hat{x}}\sigma}-\mathrm{H.c.}). 
\end{equation}
Assuming $A$ is small in Eq.~(\ref{eq:H-AJ}), the one-body wave function 
within the first-order perturbation is given by 
\begin{equation}
\Phi^{(1)}(A)=|\tilde{0}\rangle
+A\sum_{m\neq0}\frac{\langle m|J_x|\tilde{0}\rangle}{E_0-E_m}|m\rangle,
\label{eq:Phi_expansion}
\end{equation}
where $|\tilde{0}\rangle$ is the ground state of the mean-field Hamiltonian 
(${\cal H}_{\rm MF}$)
(equivalent to $\Phi_{\rm AF}$), $|m\rangle$ represents the excited states 
of ${\cal H}_{\rm MF}$,\cite{note-m} and $E_0$ and $E_m$ are the 
corresponding energies. 
To consider the perturbed term in Eq.~(\ref{eq:Phi_expansion}), we apply 
a Fourier transformation to $J_x$: 
\begin{equation}
J_x=-2t\sum_{q_x}\sin q_x\sum_{q_y,\sigma}c_{\bold{q}\sigma}^\dag 
c_{\bold{q}\sigma}. 
\label{eq:J(q)}
\end{equation}
We may replace $c_{\bold{q}\sigma}^\dag c_{\bold{q}\sigma}$ with the 
operators of Bogoliubov quasiparticles for Fermi sea, $d$-wave BCS or 
AF states, according to the choice of mean fields. 
Because Eq.~(\ref{eq:J(q)}) is already a diagonal form for the Fermi sea and 
BCS's Bogoliubov quasiparticles, $\langle m|J_x|\tilde{0}\rangle$ vanishes. 
Consequently, the normal and $d$-wave pairing states do not have a 
first-order correction in Eq.~(\ref{eq:Phi_expansion}). 
On the other hand, if we substitute AF-Bogoliubov quasiparticles in 
Eq.~(\ref{eq:J(q)}), we have
\begin{eqnarray}
J_x&=&-2t\sum_{\bold{q}\in\mathrm{MBZ}}\sin q_x\left[
(\alpha_\bold{q}^2-\beta_\bold{q}^2)(a_{\bold{q}\sigma}^\dag 
a_{\bold{q}\sigma}-a_{\bold{q+Q}\sigma}^\dag a_{\bold{q+Q}\sigma})
\right.\nonumber\\
&&\hspace{1.5cm}\left.-2\sigma\alpha_\bold{q}\beta_\bold{q}
(a_{\bold{q}\sigma}^\dag a_{\bold{q+Q}\sigma}
+a_{\bold{q+Q}\sigma}^\dag a_{\bold{q}\sigma})
\right], 
\label{eq:J(q)AF}
\end{eqnarray}
where MBZ indicates the reduced magnetic (AF) Brillouin zone. 
Owing to the interband term $a_{\bold{q+Q}\sigma}^\dag a_{\bold{q}\sigma}$, 
$\langle m|J_x|\tilde{0}\rangle$ does not vanish in this case. 
In summary, for band-insulating states such as the AF state, we need to 
add an imaginary corrective term in Eq.~(\ref{eq:Phi_expansion}) to 
the ordinary one-body part. 
\par

\begin{figure}[htbp]
\begin{center}
\vspace{0.2cm} 
\includegraphics[width=7.5cm,clip]{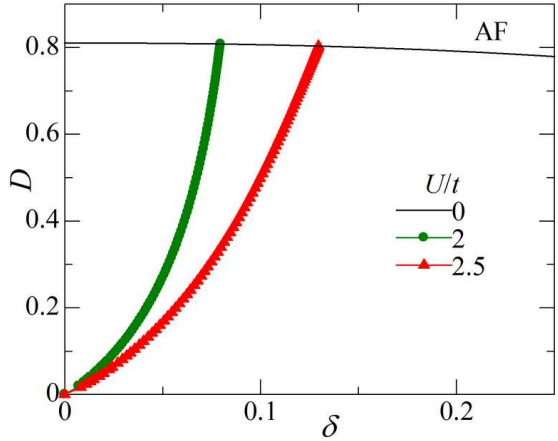} 
\end{center}
\vspace{-0.5cm} 
\caption{(Color online) 
Doping-rate dependence of Drude weight of AF state estimated using a 
mean-field theory for a couple of small values of $U/t$. 
}
\label{fig:Drude_AF_MF}
\end{figure}
%
Within the present VMC formalism, it is not easy to calculate $D$ directly 
using a wave function like Eq.~(\ref{eq:Phi_expansion}). 
We need to develop a technique to treat multiple determinants.
Instead, we can calculate $D$ of the AF state within a mean-field 
theory.\cite{SWZ} 
In Fig.~\ref{fig:Drude_AF_MF}, we show doping-rate dependence of $D$ thus 
estimated for small values of $U/t$. 
At half filling, vanishing of $D$ is realized. 
As $\delta$ increases, $D$ should linearly increases for small $\delta$'s, 
assuming that the AF state is intrinsically stable. 
However, it is known that doped AF states are unstable toward phase 
separation for the simple square lattice,\cite{doped_Mott} so that, 
actually, AF states are not realized for $\delta>0$. 
Anyway, this analysis of AF states reveals that the mechanism of reducing 
$D$ in band insulators like AF states is distinct from that in Mott 
insulators. 
\par 

\section{Attractive Hubbard model\label{sec:attractive}}
%
In this section, we study the behavior of SC weight $D_{\rm s}$ and Drude 
weight $D$ in the attractive Hubbard model ($U/t<0$), when the
configuration-dependent phase factor ${\cal P}_\theta$ and a refined 
version is applied. 
\par

Using a canonical transformation on a bipartite lattice,\cite{Dichtel,Shiba} 
one can map the physics of repulsive Hubbard model at half filling in 
magnetic fields to the physics of attractive Hubbard model.\cite{Nagaoka} 
By applying this transformation, the form of Gutzwiller projection 
${\cal P}_{\rm G}$ remains intact ($1\le g\le\infty$), but the D-H binding 
projection ${\cal P}_Q$ for $U/t>0$ is transformed to 
\begin{eqnarray}
\tilde{{\cal P}}_Q&=&\prod_j(1-\mu\tilde{Q}_j),\\
\tilde{Q}_j&=&s_j^\uparrow\prod_\tau(1-s_{j+\tau}^\downarrow)
+s_j^\downarrow\prod_\tau(1-s_{j+\tau}^\uparrow).
\label{eq:P_Q-A}
\end{eqnarray}
for $U/t<0$.\cite{Y-PTP} 
Here, $s_j^\sigma=n_{j\sigma}(1-n_{j-\sigma})$ and $\mu$ ($0\leq \mu \leq 1$) 
is a variational parameter which controls the binding between up- and 
down-spin electrons.
In a previous paper,\cite{Tamura-attractive} it was shown using VMC 
calculations that the normal state 
$\Psi_{\rm N}=\tilde{\cal P}_Q{\cal P}_{\rm G}\Phi_{\rm F}$ undergoes a 
transition between a metallic and a spin-gapped phases at 
$|U_{\rm c}|/t\sim 9$ irrespective of electron density. 
On the other hand, a homogeneous $s$-wave singlet pairing state 
$\Psi_s=\tilde{\cal P}_Q{\cal P}_{\rm G}\Phi_s$ 
exhibits a crossover in the SC properties from a BCS type to a Bose-Einstein 
condensation type at $|U_{\rm co}|/t\sim 8.7$. 
\par 

The SC weight was calculated using ${\cal P}_{\rm G}\Phi_s$\cite{Denteneer} 
and $\Psi_s=\tilde{\cal P}_Q{\cal P}_{\rm G}\Phi_s$\cite{Tamura-attractive}
as functions of $U/t$ ($<0$). 
Because these wave functions are real, resultant $D_{\rm s}$'s are 
substantially a half of the absolute values of kinetic energies (see, for 
instance, Fig.~18 in Ref.~\cite{Tamura-attractive}). 
Nevertheless, the results are broadly consistent with those of quantum Monte 
Carlo\cite{Singer} and DMFT\cite{Garg,Bauer,Toschi} calculations. 
Thus, the improvement by phase factors is expected to be small in contrast 
to the repulsive case. 
Here, we check the effectiveness of the phase factors for $U/t<0$.
\par 

\begin{table}
\caption{
Comparison of total energy [$L=12$, $A=0.5\times(2\pi/L)$] of SC state 
among cases with and without ${\cal P}_\theta$ and of correlator product 
state (CPS) for attractive interaction. 
The values in brackets indicate errors in the last digits. 
}
\begin{tabular}{c|c|c|c|c}
\hline
\multicolumn{2}{c|}{}       &\multicolumn{3}{|c}{$U/t$}\\ 
\hline
 & $\delta$ & $-5$          & $-10$         &  $-20$         \\ 
\hline\hline
 & $\ 0.0278\ $ & $\ -3.13694(5)\ $ & $\ -5.25918(9)\ $ & $\ -9.93121(6)\ $ 
\\ \cline{2-5}
 no ${\cal P}_\theta\ $ 
 & $0.0833$ & $-2.99487(6)$ & $-4.97905(6)$ & $-9.37431(5)$ \\ \cline{2-5}
 & $0.1389$ & $-2.84941(5)$ & $-4.69652(6)$ & $-8.81596(5)$ \\ \cline{2-5}
 & $0.1944$ & $-2.70042(5)$ & $-4.41163(6)$ & $-8.25631(5)$ \\ 
\hline 
 & $0.0278$ & $-3.13705(6)$ & $-5.25923(6)$ & $-9.93126(8)$ \\ 
\cline{2-5}    
 ${\cal P}_\theta$ 
 & $0.0833$ & $-2.99501(5)$ & $-4.97908(5)$ & $-9.37431(6)$ \\ \cline{2-5}
 & $0.1389$ & $-2.84955(5)$ & $-4.69659(7)$ & $-8.81601(6)$ \\ \cline{2-5}
 & $0.1944$ & $-2.70055(4)$ & $-4.41171(6)$ & $-8.25635(5)$ \\ 
\hline 
 & $0.0278$ & $-3.13717(5)$ & $-5.25942(5)$ & $-9.93149(7)$ \\ \cline{2-5}
 CPS                                          
 & $0.0833$ & $-2.99507(5)$ & $-4.97926(7)$ & $-9.37456(5)$ \\ \cline{2-5}
 & $0.1389$ & $-2.84965(5)$ & $-4.69675(7)$ & $-8.81623(5)$ \\ \cline{2-5}
 & $0.1944$ & $-2.70065(4)$ & $-4.41185(6)$ & $-8.25658(6)$ \\
\hline
\end{tabular}
\label{tb:Drude_at_L12}
\end{table}

As a first choice, we consider the same form ${\cal P}_\theta$ in 
Eq.~(\ref{eq:Ptheta}),\cite{notePthetaA} and calculate $D_s$ ($D$) with 
${\cal P}_\theta\Psi_s$ (${\cal P}_\theta\Psi_{\rm N}$). 
For a large $|U|/t$, most electrons form doublons, but ${\cal P}_\theta$ 
does not act on the hopping processes concerning isolated doublons. 
Furthermore, ${\cal P}_\theta$ does not cancel the phase $\pm 2A$ of a 
doublon hopping 
$c_{i\uparrow}^\dag c_{j\uparrow}c_{i\downarrow}^\dag c_{j\downarrow}$. 
Thus, the effect of ${\cal P}_\theta$ is expected to be limitative. 
In Table~\ref{tb:Drude_at_L12}, we compare $E/t$ between the cases with 
and without ${\cal P}_\theta$ for some values of $\delta$ and $U/t$. 
We find that $E$ (equivalently $D_{\rm s}$) decreases by applying 
${\cal P}_\theta$, but the decrement is extremely small for any $U/t$ 
and $\delta$ (orders of $10^{-5}t$-$10^{-6}t$), as expected. 
To corroborate the fact that phase factors are ineffective in this case, 
we introduce a refined phase factor in terms of the correlator product 
state (CPS).\cite{CPS1,CPS2}
The phase factor used here depends on the electron configurations of local 
five-site clusters, and has $4^5$ phase parameters to be optimized. 
As shown in lower rows of Table~\ref{tb:Drude_at_L12}, even the results 
of CPS improve the energy only slightly. 
In conclusion, a configuration-dependent phase factor is not an important 
element of $D_{\rm s}$ for $U/t<0$.
\par

\section{Summary and Discussions\label{sec:summary}}
%
In this study, we considered how to appropriately calculate the Drude and 
SC weights in the variation theory for the Hubbard model. 
We argued that existence of the imaginary part is indispensable in the wave 
function for suppressing $D$ (and $D_{\rm s}$). 
In strongly correlated regimes where Mott physics prevails, a phase 
correlation factor [${\cal P}_\theta$ in Eq.~(\ref{eq:Ptheta})], which 
depends on the local electron configuration, works successfully to estimate 
$D$ and $D_{\rm s}$ in normal and SC states, respectively. 
Thereby, at half filling, $D$ vanishes for $U>U_{\rm c}$, where 
$U_{\rm c}/t$ is the Mott transition points previously determined by other 
quantities such as doublon density and charge susceptibility 
[Figs.~\ref{fig:Drude_dope}(a) and \ref{fig:Drude_dope}(b)].  
Namely, Mott transitions are correctly described using $D$. 
Thus, the long-standing problem of Millis and Coppersmith\cite{M-C} was 
resolved. 
\par

Doping-rate dependence of $D$ and $D_{\rm s}$ for $U>U_{\rm c}$ is linear 
for $\delta\rightarrow 0$ and widely an increasing function of $\delta$ 
with some convexity [Figs.~\ref{fig:Drude_dope}(c) and 
\ref{fig:Drude_dope}(d)]. 
This behavior is consistent with what is observed experimentally for 
cuprate SC's.\cite{Uemura,Tallon}
On the other hand for weak correlations ($U<U_{\rm c}$), $D$ is finite 
at half filling, and only weakly dependent on $\delta$ especially for 
$D_{\rm s}$. 
\par

As for AF states, which is insulating for any positive $U/t$ at half 
filling in the hypercubic lattices, ${\cal P}_\theta$ is still effective 
for $U\gtrsim W$, and $D$ almost vanishes. 
However, for $U\lesssim W$ [band-insulating (Slater) regime], $D$ estimated 
using ${\cal P}_\theta\Psi_{\rm AF}$ becomes finite and smoothly approaching 
the value of noninteracting ($U=0$) limit $|E_{\rm kin}^x|/t$ as $U/t$ 
decreases. 
To correct this flaw, a perturbation theory with respect to $A$ is useful. 
We revealed that the one-body part of $\Psi$ should have a finite imaginary 
part, which is introduced by the first-order perturbation 
[Eq.~(\ref{eq:Phi_expansion})]. 
This contribution survives for the AF states, but vanishes for the normal 
and SC states. 
Thus, it can be shown $D$ vanishes for $\Phi_{\rm AF}$ at half filling 
within the mean-field theory.\cite{SWZ}
In a band insulator like $\Psi_{\rm AF}$ with $U\ll U_{\rm c}$, $D$ vanishes 
through a mechanism different from the doublon-holon binding effect in a 
Mott insulator. 
\par

For the attractive Hubbard model, we checked the effect of 
configuration-dependent phase factors in calculating $D$ ($D_{\rm s}$) 
in normal ($s$-wave SC) states.
The effect of phase factors exists, but is considerably small as compared 
to the repulsive model of a large $U/t$. 
\par

In the remainder, we make a few discussions related to the present 
subject. 
\par

(i) 
The Drude (or SC) weight is an important quantity, but, technically, $D$ 
does not seem a very suitable measure to distinguish a conductor from 
an insulator at least in the variation theory, because $D$ requires fine 
tuning of the imaginary part of the trial states according to the situations 
of individual states. 
As an alternative measure, localization length $\lambda$, which estimates 
the degree of insulating and diverges in conductive phases,\cite{xi} is 
more convenient in the sense that special tuning is not necessary for the 
trial states in appropriately computing $\lambda$.\cite{Tamura-xi} 
\par

(ii) 
As mentioned, it was shown by introducing a configuration-dependent phase 
factor like ${\cal P}_\theta$ that a staggered flux (or $d$-density wave) 
state, in which a local circular current flows in each plaquette of the 
square lattice alternately, is considerably stabilized as compared to the 
corresponding normal state in a strongly correlated Hubbard model.\cite{SF} 
Without the phase factor, we have no energy reduction. 
The physics in this case is similar to the present one; the vector potential 
of the staggered flux is almost (for $\delta=0$) or partially 
(for $\delta>0$) cancelled by ${\cal P}_\theta$. 
Similar results were also obtained for a $d$-$p$ model\cite{d-p} and 
a Bose Hubbard model\cite{Toga} with strong correlations. 
It is probable that a phase factor like ${\cal P}_\theta$ is generally 
necessary for describing current-carrying states in the regime where Mott 
physics is relevant. 
Inversely, if we cannot find an effective phase factor for a 
current-carrying state, it is probably not stabilized. 
\par

(iii) 
In this paper, we chiefly treated a fundamental phase factor 
${\cal P}_\theta$, which captures the essence of Mott physics. 
For quantitative improvement, it is intriguing to adopt refined techniques 
such as CPS discussed in Sec.~\ref{sec:attractive}. 
\par

\bigskip
\begin{acknowledgment}
\noindent
{\bf Acknowledgments}
\par\medskip
This work is supported in part by Grant-in-Aids from the Ministry of 
Education, Culture, Sports, Science and Technology, Japan. 
\end{acknowledgment}



\begin{thebibliography}{100}
\bibitem{note-dim}
Regarding nonmagnetic Mott transitions, dimensionality and connectivity of 
the lattice do not seem to make a qualitative difference at least in this 
type of theory. 
As we will mention, we confirmed that the results of Millis and Coppersmith 
basically hold also in two dimensions.
\bibitem{note-size} 
As discussed in Ref.~\citen{Miyagawa}, appreciable system-size dependence 
of $U_{\rm c}/t$ probably stems from the short-range correlation factor 
we adopt in this study. 
However, this is irrelevant to the present subject.
\bibitem{notePthetaA}
As a configuration-dependent phase factor, the canonical 
transformation\cite{Dichtel,Shiba} of ${\cal P}_\theta$ 
[Eq.~(\ref{eq:Ptheta})] seems to be suitable: 
%
\begin{equation}
\tilde{\cal P}_\theta=\exp\left[\frac{i\theta}{2}\sum_{j,\sigma}
S_j^\sigma\left(S_{j+1}^{-\sigma}-S_{j-1}^{-\sigma}\right)
\right],
\label{eq:Ptheta-A}
\end{equation} 
%
where we symmetrize the spin index. 
However, this phase factor gives unity to any electron configuration, 
so that this form is useless. 
Note that D and H are asymmetric with respect to hopping, namely, an 
electron necessarily hops from D to H. 
Therefore, the phase factor has to be asymmetric between D and H as in 
${\cal P}_\theta$ [Eq.~(\ref{eq:Ptheta})]. 
On the other hand, hopping takes place symmetrically 
($\mbox{S}^\uparrow\leftrightharpoons\mbox{S}^\downarrow$)
between singly occupied sites S$^\uparrow$ and S$^\downarrow$. 
Accordingly, the phase factor has to be symmetrized as in 
Eq.~(\ref{eq:Ptheta-A}). 
Thus, a meaningful phase projector is still ${\cal P}_\theta$ 
[Eq.~(\ref{eq:Ptheta})] for a negative $U/t$.  
In fact, by applying the cannonical transformation\cite{Dichtel,Shiba} 
the kinetic part of the Hamiltonian in a field $A$ in Eq.~(\ref{eq:H(A)}) 
is transformed to 
%
\begin{eqnarray}
{\cal H}(A)_{\rm kin}=-t\sum_{\langle i,j \rangle,\sigma} 
\left[e^{-i{\rm sgn}(\sigma)\bold{A}\cdot
(\bold{r}_i-\bold{r}_j)}c_{i\sigma}^\dag
c_{j\sigma}+{\rm H.c.}\right],
\label{eq:spinH}
\end{eqnarray} 
%
which is not a form for offerring $D$ for a charge current 
but for a spin current. Thus, the phase factor Eq.~(\ref{eq:Ptheta-A}) 
probably works for this Hamiltonian.
\end{thebibliography}

\end{document}